\documentstyle[prb,eqsecnum,aps,epsf]{revtex}
\begin{document}
\def \beq{\begin{equation}}
\def \eeq{\end{equation}}
\def \beqarr{\begin{eqnarray}}
\def \eeqarr{\end{eqnarray}}

\draft

\title{Low energy collective modes, Ginzburg-Landau theory,
and pseudogap behavior
in superconductors with long-range 
pairing interactions}

\author{Kun Yang}
\address{
National High Magnetic Field Laboratory and Department of Physics,
Florida State University, Tallahassee, Florida 32310$^*$
}
\address{and 
Condensed Matter Physics 114-36, California Institute of Technology,
Pasadena, California 91125}

\author{S. L. Sondhi}
\address{
Department of Physics,
Princeton University,
Princeton, New Jersey 08544
}

\date{\today}
\maketitle
\begin{abstract}
We study the superconducting instability in systems with long but finite
ranged,
attractive, pairing interactions. We show that such long-ranged
superconductors exhibit a new class of fluctuations in which the
internal structure of the Cooper pair wave function is soft, and
thus lead to ``pseudogap" behavior in which the actual transition
temperature is greatly depressed from its mean field value. 
These fluctuations are {\it not} phase fluctuations of the standard
superconducting order parameter, and lead to a highly unusual Ginzburg-Landau
description.
We suggest that the crossover between the BCS limit of a short-ranged
attraction and our problem is of interest in the context of 
superconductivity in the underdoped cuprates.
\end{abstract}

\section{Introduction}

Superconductivity is a subject that has attracted prolonged interest in
the condensed matter community. Much of our current understanding of this
subject is based on the standard Bardeen-Cooper-Schrieffer (BCS)\cite{bcs} 
theory, which 
has enjoyed great success when applied to conventional superconductors.
The discovery of the ``high-$T_c$'' cuprate superconductors 
opened a new chapter in
the study of superconductivity. It is now well understood that just like in
the BCS theory, electrons in cuprate superconductors form Cooper pairs.
Moreover, it has been established that the wave functions of these Cooper pairs 
have predominantly $d$-wave symmetry. However, the origin of the force that
leads to Cooper pairing in these systems is not yet understood and, 
experimentally, significant deviations from the BCS framework (which we take to
include a Fermi liquid description of the normal state) have been found
near and above the transition temperature,\cite{anderson}
especially in the underdoped 
cuprates.\cite{timusk}
Theoretically, there is a continuum of proposals for understanding these
anomalies which range from conceptually modest (if practically far reaching)
enhancements of the BCS framework (spin fluctuations,\cite{pines}
the crossover to pre-formed pairs,\cite{randreview} etc),
to the influence of various 
hypothesized quantum critical points\cite{critical} 
on to spin charge separation 
and non-Fermi liquid normal states.\cite{anderson,palee,marginal} 
Finally there 
are scenarios based on self-organized dimensional reduction (stripes) which are 
very different in spirit.\cite{emkiv}

This paper, also motivated by the physics of the cuprates, is firmly in the 
conceptually modest camp. We examine a new crossover out of the BCS corner, 
this time to a superconductor with long range pairing interactions, i.e. 
in which the range of the pairing potential $L$ is much longer than the 
coherence length ($L\gg \xi$) deduced from measurements of the (anisotropic)
gap in the quasiparticle spectrum. We remind the reader that in
the BCS theory, the phonon-mediated, retarded attraction between electrons
is often modeled as a short-range instantaneous attractive potential. As a 
matter of fact, it is often modeled as a constant attractive scattering
potential in momentum (${\bf k}$) space, which corresponds to a 
$\delta$-function attraction in real space.\cite{tinkham,schrieffer} 
This is an excellent approximation because
in conventional superconductors the coherence length $\xi$ is of order 
$\sim 1000{\rm \AA}$, while the phonon-mediated attraction is a local
(but retarded) on-site 
interaction with range of order the lattice spacing, and therefore the range is
effectively zero compared to the size of the Cooper pairs. 
In such a short range model, the elementary excitations are
quasiparticles with a (${\bf k}$ independent) excitation gap $\Delta$; 
the only collective mode below $2\Delta$ is the linear Goldstone mode, whose
energy is pushed up to the plasmon energy in real systems due to the
existence of the long-range Coulomb interaction. The classic BCS limit also involves
the assumption of weak coupling \cite{fn-strongcoup} whereupon thermally excited
quasiparticles control the largely mean-field superconducting transition 
temperature $T_c$, giving rise to a universal ratio $2\Delta/T_c\approx 3.5$.

Our interest in what happens when the range of the interaction is no longer the
lattice spacing but instead crosses the (suitably defined, see below) coherence 
length arises from four considerations. 

First, $\xi$ is much shorter in the cuprates 
than in conventional superconductors; 
in fact $\xi$ is of order five lattice spacings in the a-b plane, 
and less than one 
lattice spacing along the c-axis. 
Therefore even if it turns out that the effective 
attractive interaction that gives rise to 
pairing is of the order of a lattice spacing, 
$L$ and $\xi$ will be comparable.

Second, the physics of the pseudogap regime is most simply explained by invoking
a suppression of $T_c$ by fluctuations that are small in the BCS limit. Quite
generally the order parameter (or pair wavefunction) involves a relative
piece and a center of mass piece, and the latter 
is what enters the standard Ginzburg-Landau
theory. The simplest possibility is that the phase of the center of mass piece
fluctuates strongly and this has been suggested in the context of the cuprates
from two different standpoints, as the physics of strongly 
coupled pre-formed pairs
\cite{randreview} or that of a low superfluid density coming from a doped
Mott state.\cite{ek} 
A more novel possibility is that the relative pair wavefunction
has soft fluctuations, and as we demonstrate in this paper,
these arise in the limit of a long range attraction.

Third, it appears that the model we study here is relevant to some of the
proposed mechanisms for cuprate superconductivity. The first example is the
inter-layer pair hopping mechanism proposed by Anderson and 
co-workers.\cite{anderson,chakravarty}
In this theory, pairing is induced by coherent {\em pair hopping} along the
c-axis, provided
that {\em single electron} hopping is frustrated by the non-Fermi liquid
nature of the normal state. It was emphasized\cite{chakravarty} that 
high $T_c$
comes from the fact that the matrix element for pair hopping is {\em diagonal}
in ${\bf k}$ space, although the symmetry of the order parameter is determined
by the in-plane pairing potential. Mathematically, the pair hopping term is
equivalent to an attractive potential off-diagonal in layer index,
even though its physical origin is {\em kinetic} energy along the c-axis;
in particular, being diagonal (or a $\delta$-function) in ${\bf k}$ space, it
corresponds to a pairing potential that has an
{\em infinite} range in real space. 
A second example of this is the spin-fluctuation mechanism proposed by Pines,
Scalapino, and co-workers.\cite{pines} In this theory, the range of the effective 
pairing interaction mediated by antiferromagnetic spin fluctuations is the spin-spin
correlation length; as one reduces the doping level and
approaches the antiferromagnetic instability from
the overdoped side (where the system is more BCS-like), 
the spin-spin correlation length and therefore the range of the pairing
interaction increases; it can become very large in the underdoped region, and
{\em diverges} when approaching the antiferromagnetic instability. 

Finally, given the importance of the subject of superconductivity, we feel 
the situation we study here (which is opposite to the familiar short-range
attraction limit) is interesting in its own right. Even if it proves not to be
of particular relevance to superconductivity in cuprates, it may become relevant
elsewhere.\cite{exciton}

Our main results may be summarized as following. We find in addition to 
the usual quasiparticle excitations, the system supports 
collective modes whose energies are significantly lower than the quasiparticle
gap $\Delta$, when $L\sim \xi$; in the limit that $L\gg \xi$, the lowest energy 
collective mode gap becomes {\em much lower} than $\Delta$,
and the number of collective modes below $2\Delta$ becomes large.
At finite temperatures, due to the thermal
fluctuations of these collective modes,
the transition temperature $T_c$ becomes significantly lower than the
mean field transition temperature, $T_c^{MF}$ (which is controlled by the 
thermally excited quasiparticles and does not know about the collective modes). 
For $T_c < T < T_c^{MF}$, the electrons are paired 
(in momentum space) but there 
is no long-range phase coherence; thus the system exhibits pseudogap behavior.
Also the effective Ginzburg-Landau 
theory needs to be modified in this case; in addition to the center of mass
degree of freedom of the superconducting order parameter (or Cooper pairs),
which is the only degree of freedom taken into account in the original
Ginsburg-Landau
description, we also need to take into account the fluctuations of the
{\em relative} or {\em internal} degrees of freedom of the order parameter.
As a consequence spatial gradients couple to the internal pair structure in
non-trivial ways, and although we have not done the computation yet, we
suspect that the critical field $H_{c2}$ will be governed by a coherence
length that is distinct from the one derived from the gap.

In the main part of this paper, we will take the attitude that this is an
interesting model system to study, and work out various properties of such a
model, especially those that are different from the short-range limit of the
BCS theory. Discussion of the possible relevance of our considerations to the
cuprates will be reserved to the Summary section.

The rest of the paper is organized as following. In Section 
\ref{cooper} we revisit the Cooper problem, and keep the finite range of the
attractive potential explicit. We note some problems involved in taking
the limit of long-ranged potentials and specify the precise nature of
the limit we consider in this paper.
We find that in the limit $L\gg\xi$,\cite{note1}
there are a large number of bound state solutions, 
whose binding energies are comparable
to the ground state. These bound states correspond loosely to the collective
modes in the many-body problem. In section \ref{reduced} we study the BCS
reduced Hamiltonian with a finite range attractive force, solving for both the
mean-field ground state, and the zero momentum collective modes using the linearized
equations of motion for the order parameter. In section \ref{gl} we  
develop a Ginzburg-Landau effective theory from our model, using the 
functional integral method. We find that it is necessary to keep track of the
fluctuations of both
the internal and center of mass degrees of freedom of the superconducting order
parameter; in particular, in the limit $L\gg\xi$, $T_c$ is controlled by the
fluctuations of the internal degrees of the order parameter, and 
is much lower than $T_c^{MF}$. In Section \ref{tcreduction}
we calculate the reduction of $T_c$ from $T_c^{MF}$ due to the thermal 
fluctuations of the collective modes, using the Ginzburg-Landau theory
developed earlier. In Section \ref{summary} we discuss the possible 
relevance of our work to the cuprate superconductors and mention some
connections to other work in the literature.

\section{The Cooper Problem}
\label{cooper}

We begin by studying the Cooper problem, in which the wave function of a 
Cooper pair takes the form:
\beq
|\Psi\rangle=\sum_{0<\epsilon_k<E_c}g({\bf k})c^\dagger_{{\bf k}\uparrow}
c^\dagger_{-{\bf k}\downarrow}|\Psi_0\rangle,
\eeq
where $|\Psi_0\rangle$ is the filled Fermi sea, $\epsilon_k=v_F(k-k_F)$ is the
single electron energy measured from the Fermi level,
and
$E_c\ll E_F$ (Fermi energy)
is an energy cutoff. In phonon mediated attraction models,
$E_c$ is usually taken to be the Debye energy. In this paper we treat $E_c$ as
a parameter that can be varied in our effective model. 
The Hamiltonian (for both the Cooper problem considered here and the 
many-body problem in the BCS reduced Hamiltonian approximation
considered in the following section) takes the form
\beq
H=\sum_{{\bf k},\sigma}\epsilon_kc^\dagger_{{\bf k}\sigma}c_{{\bf k}\sigma}
-\sum_{{\bf k}{\bf k}'}V_{|{\bf k}-{\bf k}'|}c^\dagger_{{\bf k}\uparrow}
c^\dagger_{-{\bf k}\downarrow}
c_{-{\bf k}'\downarrow}c_{{\bf k}'\uparrow}.
\label{reducedeq}
\eeq
Here 
\beq
V_q={1\over A}\int{d^2{\bf r}V(r)e^{-i{\bf q}\cdot{\bf r}}}
\eeq
is the Fourier transform of an attractive potential $V(r)$;
$A$ is the area of the system.\cite{2Dnote}
It is understood that the summation of the second term of
Eq. (\ref{reducedeq}) is restricted to states
with $|\epsilon_k|, |\epsilon_{k'}| < E_c$.
For simplicity, we assume $V_q$ and $V(r)$ are positive definite, so the 
ground state for the Cooper problem has $s$-wave
symmetry, and the many-electron system is an $s$-wave superconductor. 
We will comment on the generalization to a d-wave superconductor in the
concluding section.

The Schroedinger equation that $g({\bf k})$ satisfies is
\beq
Eg({\bf k})=2\epsilon_kg({\bf k})-\sum_{{\bf k}'}V_{|{\bf k}-{\bf k}'|}
g({\bf k}').
\label{coopereq}
\eeq
where $E$ is the eigenenergy. For a bound state, we must have $E < 0$.
For a general $V_q$, the integral equation
(\ref{coopereq}) is difficult to solve. Cooper solved
the short-range limit of this problem. In this limit, $V_q=V$ is a constant
in momentum space, and therefore a $\delta$-function in real space. It was
found that there exists only one bound state solution, with $s$-wave
symmetry; for
weak coupling ($N(0)V \ll 1$, $N(0)$ being the density of states at the Fermi
energy), the energy of this state is\cite{tinkham}
\beq
E\approx -2E_ce^{-2/N(0)V}.
\label{cooperen}
\eeq
This well known dependence of the binding energy on $V$ is extremely
singular. For finite but very short-range attractive 
potentials, there may be more bound state solutions (probably in other 
angular momentum channels). However, due to the singular dependence
of the binding energy on the pairing potential, the binding energies of these
states will be much smaller than the ground state, even if the effective 
pairing potential only changes slowly from one channel to another.

Here we consider the opposite limit, in which $V(r)$ has a long but finite
range, $L$. Although the explicit form of $V(r)$ is unimportant to our
basic conclusions, for concreteness we assume it has a Gaussian form:
\beq
V(r)=V_0e^{-r^2/2L^2},
\eeq
and therefore
\beq
V_q={2\pi L^2\over A}V_0e^{-q^2L^2/2}.
\eeq

We are interested in following the evolution of the system as $L$ is
increased and in the regime where $L$ is large, $L\agt\xi$. One
possibility is to consider potentials of the ``Kac type'' familiar
from statistical mechanics, where the range is increased while keeping
the integrated strength of the potential
fixed, thereby achieving an unproblematic
thermodynamic limit. A second possibility, perhaps more appropriate to
attractive interactions generated by the system itself,\cite{fn-stab} 
is to keep
the interaction of fixed magnitude ($V_0$) but to restrict its operation to
an increasingly narrow shell around the Fermi surface. We will choose
the second course here. An easy estimate shows that we need to pick
the cutoff energy 
\beq
E_c\propto L^{-D} 
\label{ECL}
\eeq
in order to keep the potential
energy per particle finite. This also has the advantage that, by
construction, is suppresses the tendency to phase separation that would
otherwise be a complication with attractive long range 
interactions.\cite{dagotto}
There is however, a final caveat. For technical reasons, starting in
the next paragraph, we will often employ a gradient expansion in
momentum space. This is not quite consistent with the cutoff procedure
but for the results of interest we will find that the precise value of
the cutoff will enter weakly, leading us to believe that a better 
set of calculations will not alter our general scenario.

Since $V_{|{\bf k}-{\bf k}'|}$ goes to zero rapidly for $|{\bf k}-{\bf k}'| >
1/L$, if $g({\bf k})$ varies slowly on the scale of $1/L$, we may
perform a gradient expansion for $g({\bf k})$ in ${\bf k}$-space, 
in the last term
of Eq. 
(\ref{coopereq}):
\beqarr
\sum_{{\bf k}'}V_{|{\bf k}-{\bf k}'|}g({\bf k}')
&=&\sum_{{\bf k}'}V_{|{\bf k}-{\bf k}'|}\left[g({\bf k})
+\nabla_{\bf k}g({\bf k})\cdot
({\bf k}'-{\bf k})
+{1\over 2}\sum_{\mu\nu}{\partial^2 g({\bf k})\over 
\partial {k_\mu}\partial {k_\nu}}
(k'_\mu-k_\mu)(k'_\nu-k_\nu)+\cdots\right]\nonumber\\
&\approx&\left(\sum_{{\bf k}'}V_{|{\bf k}-{\bf k}'|}\right)g({\bf k})+{1\over 4}
\left(\sum_{{\bf k}'}|{\bf k}-{\bf k}'|^2V_{|{\bf k}-{\bf k}'|}\right)
\nabla_{\bf k}^2g({\bf k})\nonumber\\
&=&V_0g({\bf k})+{V_0\over 2L^2}\nabla_{\bf k}^2g({\bf k}),
\eeqarr
in which we have neglected higher gradient terms.
Thus within the gradient expansion, Eq. (\ref{coopereq}) reduces to
\beq
-{1\over 2M}\nabla_{\bf k}^2g({\bf k})+(2\epsilon_k-V_0)g({\bf k})=Eg({\bf k}),
\label{sch}
\eeq
where $M=L^2/V_0$.
This {\em differential} equation
(\ref{sch}) is identical to the Schroedinger equation of a
particle confined in an annulus (with some unknown, but calculable, boundary condition) 
with inner and outer radii $k_F$ and
$k_F+E_c/v_F$ respectively, experiencing a ``potential" $U(k)=-V_0+2\epsilon_k
=-V_0+2v_F(k-k_F)$.
It is interesting to note that in Eq. (\ref{sch}), the ``kinetic energy" term
comes from the two-body interaction in the original Hamiltonian, while the
``potential" term actually is the original kinetic energy term.

For large $L$, the ``mass" $M$ is large, and we may use WKB approximation to
analyze Eq. (\ref{sch}). The following conclusions follow straightforwardly:

\noindent
i) The ground state is in the $s$-wave channel, with energy 
\beq
E_0\approx
-V_0+ c({v_F\sqrt{V_0}\over L})^{2/3}\approx -V_0.
\eeq
Here $c$ is a constant of order 1; 
it is approximately $({3\pi\over 2\sqrt{2}})^{2/3}$ for
hard wall boundary conditions. We find the binding energy is essentially the 
depth of the attractive two-body potential $V_0$; 
this dependence is much less singular than the short-range limit
defined above. Also the cutoff $E_c$ does not enter explicitly.\cite{cutoffnote}
At this point we define a ``coherence length" 
\beq
\xi=2v_F/
\pi |E_0|
\approx 2v_F/\pi V_0,
\label{coopercoher}
\eeq
in analogy to the BCS coherence length at zero 
temperature:\cite{tinkham,schrieffer} $\xi_0=v_F/(\pi\Delta(0))$, where 
$\Delta(0)$ is the quasiparticle gap at zero temperature.
What we have in mind is that 
the binding energy in the Cooper problem will be the same as twice
the quasiparticle gap in the BCS problem.
This is {\em not} the case in short-range models;
however, as we will see later, it is indeed true in the present
case. We need to emphasize however, that $\xi$ is {\em not} the size of the
Cooper pair wave function $\ell$ in the two-body problem studied in this
section (this is the case even in short-range models);
the latter can be estimated easily from Eq.
(\ref{sch}). The size of the ground state wave function in momentum space for 
the Hamiltonian (\ref{sch}) is
\beq
\Delta k \sim \left({1\over v_FM}\right)^{1/3}
=\left({V_0\over v_FL^2}\right)^{1/3},
\eeq
thus 
$\ell\sim 1/\Delta k \sim (v_FL^2/V_0)^{1/3}$; it actually increases with $L$; 
however the sublinear dependence means for large $L$ we do have $\ell\ll L$.
This points to some ambiguities in what is meant by the coherence length
in our problem 
and presumably spatial gradients may be governed by
a different coherence length than the gap coherence length. 
We note
that this possibility has been raised on completely different grounds
in Ref. \onlinecite{palee}.

In order for
the gradient expansion performed above to be valid, we must have 
$\Delta k\agt 1/L$, i.e., $g({\bf k})$
of the ground state
varying slowly over the scale $1/L$, which is the range of $V_q$.
This leads to the condition
\beq
L\agt 2v_F/\pi V_0\approx\xi,
\label{condition}
\eeq
as advertised earlier. It turns out that throughout this paper
the gradient expansion in momentum 
space is the key technique that allows explicit calculations of various
physical quantities to be made, and the above condition defines the range of
its validity.
For a given potential with fixed $V_0$ and $L$, one may also
interpolate between
the short-range and long-range limits by varying $v_F$. In the following we
assume Eq. (\ref{condition}) is satisfied, and pay particular attention to the
limit $L\gg\xi$.

\noindent
ii) The number of bound states in the $s$-wave ($l=0$) channel is
$N_0\approx {\sqrt{2}L\over 3\xi}$.

\noindent
iii) The largest angular momentum channel that supports a bound state has
angular momentum $l_{max}\approx \sqrt{2} k_FL$.

\noindent
iv) The total number of bound states is
$N_{tot}=\sum_lN_l\approx {\pi\over 6}k_FL^2/\xi$.

\noindent
v) The energy spacing between the ground state and the first excited state is
$\Delta E\approx {V_0\over 2k_F^2L^2} \ll |E_0|$, i.e., it is much lower
than the binding energy of the ground state itself, in contrast to 
the short range case.

The gradient expansion does not apply to all bound states, even if $L\gg\xi$.
One can show that it only applies to states with energy measured from the
ground state (or bottom of the potential well)
$\Delta E \alt V_0\xi^2/L^2$. Nevertheless the estimate of number of 
bound states in various channels should be qualitatively correct.

These large number of low-energy bound states correspond loosely to the 
low-energy collective modes in the many body problem studied in the following
sections, and they lead to a significant reduction of $T_c$ from its mean-field
value $T_c^{MF}$ (whose scale is set by the gap which is the same as the Cooper
pair binding energy here, $T_c^{MF}\sim \Delta \sim V_0$). 
This may be understood 
heuristically in the following way. When the temperature $T$ is reduced to 
$T_c^{MF}$, electrons start to form Cooper pairs. In the short-range BCS model,
there is only one way to form a Cooper pair, thus all pairs condense into
one state immediately and long-range coherence is established. In the model
studied here however, there are many different ways to form Cooper pairs and
their binding energies are comparable; thus the system needs to go to much
lower temperature for all the Cooper pairs to condense into the ground state,
hence $T_c\ll T_c^{MF}$. For $T_c < T < T_c^{MF}$, the electrons are paired
(and therefore gapped),
but there is no long-range superconducting coherence, and the system
exhibit pseudogap behavior.

\section{Mean field theory and collective modes in the BCS reduced Hamiltonian}
\label{reduced}

In this section we study the BCS reduced Hamiltonian, Eq. (\ref{reducedeq}).
Instead of solving the two-body problem in the above section, here we study
the many-body problem. The difference between Eq. (\ref{reducedeq}) and the
full many-body problem is that in Eq. (\ref{reducedeq}), only pairs of 
electrons with opposite spin and total momentum zero scatter each other.
(we return to the full problem, i.e. when the sum over the interacting 
momenta is 
constrained only by momentum conservation in the next section.)
This leads to a very special property,
that an unpaired electron never gets scattered, and thus the number of 
unpaired electron (0 or 1) for any ${\bf k}$ and spin orientation
is a conserved 
quantity. The ground state for this Hamiltonian
is in the subspace in which no such unpaired 
electron exists, namely the pair of single electron states 
${\bf k}\uparrow$ and $-{\bf k}\downarrow$ are either both occupied or both
empty. In this subspace, one may map the problem onto a quantum pseudospin 
problem.\cite{anderson57} We introduce a pseudospin-1/2 operator for each
${\bf k}$:
\beqarr
S_{\bf k}^z&=&(c^\dagger_{{\bf k}\uparrow}c_{{\bf k}\uparrow}
+c^\dagger_{-{\bf k}\downarrow}c_{-{\bf k}\downarrow}-1)/2,
\nonumber\\
S_{\bf k}^+&=&c^\dagger_{{\bf k}\uparrow}c^\dagger_{-{\bf k}\downarrow},
\nonumber\\
S_{\bf k}^-&=&c_{-{\bf k}\downarrow}c_{{\bf k}\uparrow}.
\eeqarr
In this mapping, the pseudospin is pointing up when a pair of single electron 
states are occupied, and down when they are empty; pair creation/annihilation
operators map onto spin-flip operators.
The Hamiltonian now takes the form (up to a constant):
\beq
H=\sum_{{\bf k}}2\epsilon_{\bf k}S_{\bf k}^z
-{1\over 2}\sum_{{\bf k}{\bf k}'}V_{|{\bf k}-{\bf k}'|}(
S_{\bf k}^+S_{{\bf k}'}^-+S_{{\bf k}'}^+S_{\bf k}^-)
=\sum_{{\bf k}}2\epsilon_{\bf k}S_{\bf k}^z
-\sum_{{\bf k}{\bf k}'}V_{|{\bf k}-{\bf k}'|}(
S_{\bf k}^xS_{{\bf k}'}^x+S_{{\bf k}'}^yS_{\bf k}^y),
\eeq
i.e., the kinetic energy maps onto a Zeeman field that depends linearly on
$\epsilon_{\bf k}$ (and changes sign at the Fermi surface), which couples to
the $z$ component of the pseudospins; the pairing interaction maps onto a
{\em ferromagnetic} coupling among the $xy$ components of the pseudospins.

In the mean field approximation, one replaces the ferromagnetic coupling among 
the $xy$ components of the spins by an average field:
\beq
H_{MF}=-\sum_{{\bf k}}{\bf B}_{\bf k}^0\cdot{\bf S_k},
\eeq
where ${\bf B}_{\bf k}^0$ satisfies the self-consistency equation:
\beq
{\bf B}_{\bf k}^0=-2\epsilon_{\bf k}\hat{z}+2\sum_{{\bf k}'}V_{|{\bf k}-{\bf k}'|}
\langle{\bf S}_{{\bf_k}'}^\bot\rangle.
\label{mf}
\eeq
Here ${\bf S}_{\bf_k}^\bot$ stands for the $xy$ components of 
${\bf S_k}$. In the ground state, ${\bf S_k}$ points in the direction of
${\bf B}_{\bf k}^0$; the elementary excitations are single pseudospin flips,
which corresponds to a pair of Bogliubov quasiparticles with opposite 
momenta and (real) spin.\cite{note2}

In the short-range limit $L\rightarrow 0$, the range of 
$V_{|{\bf k}-{\bf k}'|}$, $1/L$, {\em diverges} in ${\bf k}$ space.
Thus all the pseudospins are equally coupled, no matter how far away they are
in ${\bf k}$ space. In this limit, the mean field approximation becomes exact;
the only collective mode is a zero mode corresponds to the global rotation
of all the pseudospins along the $z$-axis, reflecting the broken XY symmetry
of the ground state.\cite{anderson57}
All other excitations in this subspace may be described by pseudospin flips,
or pairs of quasiparticles.

The situation, however, becomes quite different, when $L$ becomes large. In
this case $V_{|{\bf k}-{\bf k}'|}$ becomes {\em short-ranged}
in ${\bf k}$ space;
one may actually divide the ${\bf k}$ space into blocks of size ${1\over L}
\times {1\over L}$; within each block all the spins are strongly coupled and
are effectively locked into a very big single spin; for different blocks,
however, couplings are restricted to neighboring blocks. In such an $XY$ 
ferromagnet with ``short range" couplings among the blocked spins, there are
(pseudo)spin-wave 
like collective excitations in addition to single pseudospin flips,
whose energies can become significantly lower than pseudospin flips for 
large $L$, as the number of blocks increases as $\sim L^2$.

We begin by analyzing the mean field equation, Eq. (\ref{mf}), at
zero temperature. Without losing
generality, we may assume ${\bf B}_{\bf k}^0$ is in the $x-z$ plane:
\beq
{\bf B}_{\bf k}^0=-2\epsilon_{\bf k}\hat{z}+2\Delta_{\bf k}\hat{x},
\eeq
and the self-consistency equation for $\Delta_{\bf k}$ is
\beq
\Delta_{\bf k}=\sum_{{\bf k}'}V_{|{\bf k}-{\bf k}'|}
\langle{\bf S}_{\bf_k}^x\rangle
={1\over 2}\sum_{{\bf k}'}{V_{|{\bf k}-{\bf k}'|}\Delta_{{\bf k}'}\over
\sqrt{\epsilon_{{\bf k}'}^2+\Delta_{{\bf k}'}^2}},
\label{self}
\eeq
which is the familiar BCS gap equation.\cite{tinkham,schrieffer}
In the 
limit $L\rightarrow 0$, $V_{|{\bf k}-{\bf k}'|}=V$ becomes independent of
$|{\bf k}-{\bf k}'|$, so does $\Delta_{\bf k}=\Delta$,
and Eq. (\ref{self}) reduces to 
\beq
1={1\over 2}\sum_{{\bf k}'} {V\over
\sqrt{\epsilon_{{\bf k}'}^2+\Delta^2}}.
\eeq
In the weak coupling
limit, its solution is $\Delta=2E_ce^{-1/N(0)V}$. This gap is much larger than
the binding energy in the Cooper problem, Eq. (\ref{cooperen}), due to the 
factor of two difference in the exponential.

The situation becomes very different in the opposite limit, that
$L$ becomes large. In this case $V_{|{\bf k}-{\bf k}'|}$ becomes short-ranged
in ${\bf k}$ space; thus on the right hand side of Eq. (\ref{self}),
we may replace $\Delta_{{\bf k}'}$ and $\epsilon_{{\bf k}'}$ by
$\Delta_{{\bf k}}$ and $\epsilon_{{\bf k}}$ to first approximation,
provided that they vary slowly on the scale of $1/L$. In this approximation
one finds in the limit $L\rightarrow\infty$,
\beq
\Delta_{{\bf k}}=\sqrt{{V_0^2\over 4}-\epsilon_{{\bf k}}^2},
\label{gap}
\eeq
for $|\epsilon_{{\bf k}}| < V_0/2$, and $\Delta_{{\bf k}}=0$ otherwise.
For large but finite $L$, $\Delta_{{\bf k}}$ is nonzero but exponentially
small for $|\epsilon_{{\bf k}}| > V_0/2$.
Thus $\Delta_{{\bf k}}$ is ${\bf k}$ {\em dependent}, and reaches its maximum
for $k=k_F$, in which case
\beq
\Delta_{k_F}={V_0\over 2}.
\eeq
We find the quasiparticle gap and the binding energy are indeed set by the
same energy scale in the Cooper problem, as advertised earlier; in
particular, the gap for creating a pair of 
quasiparticles on the Fermi surface, $2\Delta_{k_F}=V_0$, 
is exactly the Cooper pair
binding energy, in the limit $L\rightarrow \infty$. Following standard
convention,\cite{tinkham} we introduce the coherence length
\beq
\xi={v_F\over \pi\Delta_{k_F}}\approx {2v_F\over \pi V_0}.
\eeq
This definition indeed matches that of $\xi$ in the Cooper problem Eq. 
\ref{coopercoher}, in the
limit that $L$ is large. And one can easily show that the large $L$ 
approximation is valid for
\beq
L\agt\xi.
\eeq
Within the mean field theory, 
the elementary excitations are the Bogliubov quasiparticles, with the standard
spectrum
\beq
E_k=\sqrt{\epsilon_k^2+\Delta_k^2}.
\eeq

At finite temperature, the self-consistent mean field equation becomes
\beq
\Delta_{\bf k}
={1\over 2}\sum_{{\bf k}'}{V_{|{\bf k}-{\bf k}'|}\Delta_{{\bf k}'}\over
E_{\bf k}}
\tanh{\beta E_{\bf k}\over 2},
\label{tself}
\eeq
where $\beta={1\over T}$ is the
inverse temperature (we set the Boltzmann constant $k_B$ to be 1).
We can use the above equation to determine the mean-field transition temperature
$T_c^{MF}$.
In the $L\rightarrow\infty$ limit, it is particularly simple;
in this case Eq. (\ref{tself})  
reduces to 
\beq
1={V_0\over 2E_{\bf k}}\tanh{\beta E_{\bf k}\over 2}.
\label{mftc}
\eeq
Using the facts that as $T\rightarrow T_c^{MF}$, 
$\Delta_{\bf k}\rightarrow 0$ and
$E_{\bf k}\rightarrow \epsilon_{\bf k}$,
and focusing on $|{\bf k}|\sim
k_F$
where the pairing instability is the strongest,
we find 
\beq
T_c^{MF}={V_0\over 4}
\eeq
in this limit.
Thus the ratio $2\Delta_{k_F}/T_c^{MF}=4$, which is slightly bigger than the BCS
ratio of $3.5$.

There are a couple of new features of the mean-field solution in the large
$L$ regime, that deserve some further discussion. (i) Unlike the case of 
short-range attraction originally considered by BCS, where the gap $\Delta$ is
a constant in momentum space at all temperatures (within the cutoff where the
pairing interaction is nonzero), 
here $\Delta_{\bf k}$ has strong momentum dependence, being maximum at
the Fermi surface and decreasing as one moves away from it. At $T=0$, the 
momentum dependence Eq. (\ref{gap}) 
is such that the quasiparticle dispersion is
essentially {\it flat} near the Fermi surface, for sufficiently large $L$.
(ii) The temperature $T$ not only affects the overall scale of the gap, but 
also its momentum dependence. Taken at face value, one would conclude from
Eq. (\ref{mftc}) that $T_c^{MF}$ depends on $\epsilon_{\bf k}$ in the following
way:
\beq
T_c^{MF}(\epsilon_{\bf k})=
T_c^{MF}{2\epsilon_{\bf k}/V_0\over \tanh^{-1}(2\epsilon_{\bf k}/V_0)}
\le T_c^{MF}.
\eeq
What this really means, of course, is that the gap $\Delta_{\bf k}$ remains
exponentially small for $T_c^{MF}(\epsilon_{\bf k}) \alt T < T_c^{MF}$.
Thus in some sense $T_c^{MF}$ is a ``local" property in the momentum space,
increasing monotonically, and
scaling roughly with the size of local gap $\Delta_{\bf k}(T=0)$ unless
$\Delta_{\bf k}(T=0)$ is very small.

In the following we go beyond mean field theory and study the collective
excitations of the system at $T=0$. 
To do that, we study the equations of motion for
${\bf S_k}$\cite{anderson57}:
\beq
{d{\bf S_k}\over dt}=-i[{\bf S_k}, H]={\bf S_k}\times {\bf B_k},
\eeq
where
\beq
{\bf B_k}=-2\epsilon_{\bf k}\hat{z}+2\sum_{{\bf k}'}V_{|{\bf k}-{\bf k}'|}
{\bf S}_{{\bf_k}'}^\bot.
\eeq
Write ${\bf B_k}={\bf B}^0_{\bf k}+\delta{\bf B_k}$ and 
${\bf S_k}={\bf S}^0_{\bf k}+\delta{\bf S_k}$, where 
${\bf S}^0_{\bf k}$ is the expectation value of ${\bf S}_{\bf k}$ in the
mean field ground state, and linearizing the 
equation of motion by neglecting terms proportional to $\delta{\bf S_k}\times
\delta{\bf B_k}$, we obtain
\beq
{d\delta{\bf S_k}\over dt}=\delta{\bf S_k}\times {\bf B}^0_{\bf k}
+{\bf S}^0_{\bf k}\times\delta{\bf B_k}.
\label{motioneq}
\eeq
$\delta{\bf S_k}$ should be perpendicular to ${\bf S}^0_{\bf k}$. If 
${\bf S}^0_{\bf k}$ is in the $x-z$ plane, we may assume
\beq
\delta{\bf S_k}=\delta S_{\bf k}^y\hat{y}+\delta S_{\bf k}^{\parallel}
\hat{e}_{\bf k},
\eeq
where $\hat{e}_{\bf k}$ is in the $x-z$ plane, but perpendicular to 
${\bf S}^0_{\bf k}$.
Thus Eq. (\ref{motioneq}) reduces to
\beqarr
{d\delta S_{\bf k}^y\over dt}&=&{1\over 2}\delta B_{\bf k}^x\cos\theta_{\bf k}
-B^0_{\bf k}\delta S_{\bf k}^{\parallel},\\
{d\delta S_{\bf k}^\parallel\over dt}&=& B^0_{\bf k}\delta S_{\bf k}^y
-{1\over 2}\delta B^y_{\bf k}.
\eeqarr
Here $\theta_{\bf k}$ is the angle between ${\bf B}^0_{\bf k}$ and the $\hat{z}$
direction, and
\beqarr
\delta B_{\bf k}^x&=&2\sum_{{\bf k}'}V_{|{\bf k}-{\bf k}'|}\delta 
S^\parallel_{{\bf k}'}\cos\theta_{{\bf k}'},\\
\delta B_{\bf k}^y&=&2\sum_{{\bf k}'}V_{|{\bf k}-{\bf k}'|}\delta 
S^y_{{\bf k}'}.
\eeqarr
Thus for a mode with frequency $\omega$ and $\delta S_{\bf k}^y\propto 
\phi_{\bf k}$, it must satisfy 
\beqarr
-\omega^2\phi_{\bf k}&=&-(B^0_{\bf k})^2\phi_{\bf k}+B^0_{\bf k}\sum_{{\bf k}'}
V_{|{\bf k}-{\bf k}'|}\phi_{{\bf k}'}
+\cos\theta_{\bf k}\sum_{{\bf k}'}
V_{|{\bf k}-{\bf k}'|}B^0_{{\bf k}'}\cos\theta_{{\bf k}'}
\phi_{{\bf k}'}\nonumber\\
&-&\cos\theta_{\bf k}\sum_{{\bf k}'{\bf k}"}
V_{|{\bf k}-{\bf k}'|}V_{|{\bf k}'-{\bf k}"|}\cos\theta_{{\bf k}'}
\phi_{{\bf k}"}.
\eeqarr
Performing a gradient expansion similar in spirit to the previous section, we
obtain
\beqarr
\omega^2\phi_{\bf k}&=&\tilde{U}_{\bf k}\phi_{\bf k}
-{1\over 4}B^0_{\bf k}\sum_{{\bf k}'}|{\bf k}-{\bf k}'|^2
V_{|{\bf k}-{\bf k}'|}\nabla_{\bf k}^2\phi_{\bf k}
-{\cos\theta_{\bf k}\over 4}
(\sum_{{\bf k}'}|{\bf k}-{\bf k}'|^2V_{|{\bf k}-{\bf k}'|})
\nabla_{\bf k}^2(\cos\theta_{\bf k}B^0_{\bf k}\phi_{\bf k})\nonumber\\
&+&\cos\theta_{\bf k}[\sum_{{\bf k}'{\bf k}"}
V_{|{\bf k}-{\bf k}'|}V_{|{\bf k}'-{\bf k}"|}\cos\theta_{{\bf k}'}({\bf k}"
-{\bf k})]\cdot\nabla_{\bf k}\phi_{\bf k}\nonumber\\
&+&{1\over 2}
\cos\theta_{\bf k}[\sum_{{\bf k}'{\bf k}"}
V_{|{\bf k}-{\bf k}'|}V_{|{\bf k}'-{\bf k}"|}\cos\theta_{{\bf k}'}(k_\mu"
-k_\mu)(k_\nu"-k_\nu)]\partial_\mu\partial_\nu\phi_{\bf k}+\cdots,
\label{modeseq}
\eeqarr
where the ellipses stand for terms 
that involve higher gradients which we neglect.
Here 
\beqarr
\tilde{U}_{\bf k}
&=&(B^0_{\bf k})^2-B^0_{\bf k}\sum_{{\bf k}'}V_{|{\bf k}-{\bf k}'|}
-\cos^2\theta_{\bf k}(B^0_{\bf k}\sum_{{\bf k}'}V_{|{\bf k}-{\bf k}'
|}-\sum_{{\bf k}'{\bf k}"}
V_{|{\bf k}-{\bf k}'|}V_{|{\bf k}'-{\bf k}"|})
-{V_0^2\over 2L^2}\nabla_{\bf k}^2\cos\theta_k
\nonumber\\
&=&(B^0_{\bf k}-\cos^2\theta_{\bf k} V_0)(B^0_{\bf k}-V_0)
-{V_0^2\over 2L^2}\nabla_{\bf k}^2\cos\theta_k.
\eeqarr
In the limit that $L$ becomes very large, one finds 
$\tilde{U}_{\bf k}=0$ for $|\epsilon_{\bf k}| < V_0$, and
$\tilde{U}_{\bf k}=(\epsilon_{\bf k}-V_0)^2$ otherwise.

For low energy modes, we expect $\phi_{\bf k}$ to be centered near $k=k_F$.
Thus in a somewhat crude approximation, we may neglect the gradient terms
in Eq. (\ref{modeseq}) proportional to $\cos\theta_{\bf k}$, as 
$\cos\theta_{\bf k}\rightarrow 0$ as $k\rightarrow k_F$; and assume
$\tilde{U}_{\bf k}$ to become infinite for $|\epsilon_{\bf k}| > V_0$. 
Within this approximation (which should give qualitatively correct results),
we obtain the collective mode frequencies to be
\beq
\omega_{mn}={V_0\over \sqrt{2}}
\sqrt{{m^2\over k_F^2L^2}+{\pi^4n^2\over (L/\xi)^2}},
\eeq
where $m$ and $n$ are ``momentum"-like quantum numbers along and perpendicular
to the Fermi surface respectively. So the mode frequency is linear in these
``momenta", as one would expect for the spin wave spectrum in
an XY ferromagnet. It is clear that the
energies of these collective modes become much lower than those of single
pseudospin flips (corresponding to quasiparticle excitations), when $L$ is
large. Since our discussion in this section is restricted to the BCS reduced
Hamiltonian in which only pairs with zero total momentum interact with each 
other, these modes are exciton-like collective modes at zero momentum; they
will acquire the usual dispersion of exciton modes at finite momentum.
The term ``exciton'' indicates that the zero momentum excitation consists
of a pair of quasiparticles bound into a pair state {\it different} from
that of the condensate which is why the state lies in the gap between the
ground state and the two quasiparticle continuum. The availability of such
states (as we already saw in the previous section)
is due to the long range of the interaction.

\section{Ginzburg-Landau theory and electromagnetic response}
\label{gl}

In this section we use the functional integral formalism to derive the effective
Ginzburg-Landau theory near their $T_c^{MF}$ for superconductors with finite
range attractive interactions, and use it to derive their analog of the London
equation.

Let us consider the Hamiltonian
\beq
\hat{H}=\hat{T}+\hat{V}
=\sum_{k\sigma}(\epsilon_k-\mu)c^{\dagger}_{{\bf k}\sigma}
c_{{\bf k}\sigma}
-\int
{d{\bf x}{d{\bf x}'}V(|{\bf x}-{\bf x}'}|)\Psi_\uparrow^\dagger({\bf x})
\Psi_\downarrow^\dagger({\bf x}')\Psi_\downarrow({\bf x}')
\Psi_\uparrow({\bf x}).
\eeq
Here $V(x) \ge 0$ represents an attractive interaction. 
We will primarily be interested in singlet pairing in the ground state; 
so we neglect
interactions between electrons with the same spin.\cite{spinnote} 
While the interaction is written in a pairwise form, it is understood that
a cutoff exists in momentum space so that only electrons that are close enough
to the Fermi surface interact with each other, in the same manner as in 
previous sections. The difference here is that the interaction is no longer
restricted to pairs of electrons with zero total momentum; thus the above 
Hamiltonian represents the full many-body problem, albeit with the parallel
spin interaction ignored.   

One may also describe the system using an Euclidean action in terms of
Grassman variables:
\beq
S[\Psi, \overline{\Psi}]
=S_0[\Psi, \overline{\Psi}]-\int_0^{\beta}{d\tau}\int
{d{\bf x}{d{\bf x}'}V(|{\bf x}-{\bf x}'}|)\overline{\Psi}_\uparrow({\bf x},\tau)
\overline{\Psi}_\downarrow({\bf x}',\tau)\Psi_\downarrow({\bf x}',\tau)
\Psi_\uparrow({\bf x},\tau),
\eeq
where $S_0$ is the action for free electrons, and $\tau$ is the imaginary time.
The partition function is
\beq
Z=\int{D\overline{\Psi}D\Psi}e^{-S[\Psi, \overline{\Psi}]}.
\eeq
We now decouple the quartic term in $S$ by introducing a pair of
Hubbard-Stratonovich
fields $\Delta({\bf x}, {\bf x}', \tau)$ and
$\overline{\Delta}({\bf x}, {\bf x}', \tau)$, which will become the
superconducting order parameter in the sequel:
\beqarr
S[\Psi, \overline{\Psi}, \Delta, \overline{\Delta}]
&=&S_0-\int_0^{\beta}{d\tau}\int
{d{\bf x}d{\bf x}'}[\Delta({\bf x}, {\bf x}', \tau)
\overline{\Psi}_\uparrow({\bf x},\tau)
\overline{\Psi}_\downarrow({\bf x}',\tau)
+\overline{\Delta}({\bf x}, {\bf x}', \tau)
\Psi_\downarrow({\bf x}',\tau)
\Psi_\uparrow({\bf x},\tau)]\nonumber\\
&+&\int_0^{\beta}{d\tau}\int
{d{\bf x}d{\bf x}'}{|\Delta({\bf x}, {\bf x}', \tau)|^2\over
V(|{\bf x}-{\bf x}'|)}.
\eeqarr
With this decoupling, the fermionic action becomes quadratic, and can be
integrated out, after which we obtain an effective action in terms of the
order parameter $\Delta({\bf x}, {\bf x}', \tau)$:
\beq
S_e[\Delta, \overline{\Delta}]
=\int_0^{\beta}{d\tau d{\bf x}d{\bf x}'}
{|\Delta({\bf x}, {\bf x}', \tau)|^2\over
V(|{\bf x}-{\bf x}'|)}-\log Z[\Delta, \overline{\Delta}],
\label{action}
\eeq
where
\beq
Z[\Delta, \overline{\Delta}]
=\int{D\overline{\Psi}D\Psi}e^{-S_0[\Psi, \overline{\Psi}]
+\int_0^{\beta}{d\tau d{\bf x}d{\bf x}'}[\Delta({\bf x}, {\bf x}', \tau)
\overline{\Psi}_\uparrow({\bf x},\tau)
\overline{\Psi}_\downarrow({\bf x}',\tau)
+\overline{\Delta}({\bf x}, {\bf x}', \tau)
\Psi_\downarrow({\bf x}',\tau)
\Psi_\uparrow({\bf x},\tau)]}.
\eeq
 
The mean-field solution corresponds to the saddle point of 
$S_e[\Delta, \overline{\Delta}]$:
\beq
{\delta S_e[\Delta, \overline{\Delta}]\over \delta\overline{\Delta}
({\bf x}, {\bf x}')}\left|_{\Delta=\Delta_s}
={\Delta_s({\bf x}, {\bf x}')\over V({\bf x}-{\bf x}')}
-{\delta \log Z[\Delta, \overline{\Delta}]\over \delta\overline{\Delta}
({\bf x}, {\bf x}')}\right |_{\Delta=\Delta_s}
={\Delta_s({\bf x}, {\bf x}')\over V({\bf x}-{\bf x}')}
-\langle\Psi_\downarrow({\bf x}')\Psi_\uparrow({\bf x})\rangle_{\Delta_s}
=0,
\label{saddle}
\eeq
where $\langle\rangle_{\Delta_s}$ 
stands for quantum and thermal averaging in the presence
of the pairing field $\Delta_s({\bf x}, {\bf x}')$. Here we have assumed a 
static saddle point so that $\Delta_s$ has no $\tau$ dependence. Further
assuming that at the saddle point, $\Delta_s({\bf x}, {\bf x}')=
\Delta_s({\bf x}-{\bf x}')$
is translationally invariant, it is easy to show that Eq. (\ref{saddle})
is equivalent to Eq. (\ref{self}),
upon the identification
\beq
\Delta_{\bf k}={1\over A}\int{d{\bf x}}e^{-i{\bf k}\cdot{\bf x}}\Delta
_s({\bf x}).
\eeq

The functional integral formalism can be used to derive the effective
Ginzburg-Landau free energy, in the vicinity of $T_c^{MF}$. This has been 
done for the short-range attractive interactions.\cite{popov}
Here we use it to derive the appropriate Ginzburg-Landau free energy for
finite range attractive interactions. 

Our starting point is the effective action, Eq. (\ref{action}).
Near $T_c^{MF}$, we may make two simplifications: i) We may neglect the
$\tau$ dependence of $\Delta$ as we expect the thermal fluctuations to 
dominate the quantum fluctuations; 
ii) We may expand $S_e$ in powers of 
$\Delta$. The quadratic terms take the form
\beq
S^{(2)}_e[\Delta, \overline{\Delta}]
=\beta\int{d{\bf x}d{\bf y}}
{|\Delta({\bf x}, {\bf y})|^2\over
V(|{\bf x}-{\bf y}|)}
-\int d{\bf x}_1d{\bf y}_1d{\bf x}_2d{\bf y}_2Q({\bf x}_1, {\bf y}_1;
{\bf x}_2, {\bf y}_2) \Delta({\bf x}_1, {\bf y}_1)
\overline{\Delta}({\bf x}_2, {\bf y}_2),
\label{s2}
\eeq
where
\beqarr
&Q&({\bf x}_1, {\bf y}_1;
{\bf x}_2, {\bf y}_2) = 
{\delta^2\log Z\over \delta\Delta({\bf x}_1, {\bf y}_1)
\delta \overline{\Delta}({\bf x}_2, {\bf y}_2)}\vert_{\Delta=0}\nonumber\\
&=&\int_0^\beta d\tau_1\int_0^\beta d\tau_2\langle
\Psi_\downarrow({\bf y}_2, \tau_2)\Psi_\uparrow({\bf x}_2, \tau_2)
\overline{\Psi}_\uparrow({\bf x}_1, \tau_1)
\overline{\Psi}_\downarrow({\bf y}_1, \tau_1)\rangle_c
\nonumber\\
&=& \int_0^\beta d\tau_1\int_0^\beta d\tau_2
G_0({\bf y}_2 - {\bf y}_1; \tau_2 -\tau_1)
G_0({\bf x}_2 - {\bf x}_1; \tau_2 -\tau_1)\nonumber\\
&=&\sum_{i\omega_n}G_0({\bf y}_2-{\bf y}_1; i\omega_n)
G_0({\bf x}_2 - {\bf x}_1; 
-i\omega_n).
\eeqarr
Here $G_0$ is the {\em normal state} (non-interacting)
single electron Green's function,
$\omega_n$'s are fermion Masubara frequencies, and $\langle\rangle_c$
stands for connected contractions in the average.

We now introduce ``center of mass" and ``relative" coordinates for the
order parameter $\Delta({\bf x}, {\bf y})$:
\beq
{\bf R}=({\bf x} + {\bf y})/2, \hskip 1cm {\bf r}={\bf y} - {\bf x},
\eeq
and Fourier transform with respect to the relative coordinate ${\bf r}$:
\beq
\Delta({\bf R}, {\bf k}) = \int{d{\bf r}} e^{-i{\bf k}\cdot {\bf r}}
\Delta({\bf R}-{\bf r}/2, {\bf R}+{\bf r}/2).
\eeq
In a uniform (${\bf R}$ independent) configuration, $\Delta({\bf R}, {\bf k})$
becomes $\Delta_{\bf k}$.

In terms of $\Delta({\bf R}, {\bf k})$, the first term of Eq. (\ref{s2})
becomes
\beqarr
&\beta&\int{d{\bf x}d{\bf y}}
{|\Delta({\bf x}, {\bf y})|^2\over
V(|{\bf x}-{\bf y}|)}
=\beta\int{d{\bf R}}\int{d{\bf r}} 
{|\Delta({\bf R}-{{\bf r}\over 2}, {\bf R}+{{\bf r}\over 2})|^2\over V(r)}
\nonumber\\
&=&{\beta\over A^2}\int{d{\bf R}}\sum_{{\bf k}_1{\bf k}_2}
\Delta({\bf R}, {\bf k}_1)\overline{\Delta}({\bf R}, {\bf k}_2)
F(|{\bf k}_1-{\bf k}_2|),
\eeqarr
where 
\beq
F(|{\bf k}_1-{\bf k}_2|)=\int{d{\bf r}}e^{i({\bf k}_1-{\bf k}_2)
\cdot{\bf r}}/V(r).
\eeq
This expression is problematic as it stands, for $1/V(r)$ does not, in
a strict sense, possess a Fourier transform in a truly infinite system. 
However, it can be defined in a large but finite-size system with periodic
boundary conditions; and as we will see in
a few lines, this is only an intermediate expression and can be regulated
as such and the final answer will be sensible even as the regulator is
removed. It is easy to see that such a regulated $F(k)$ must be peaked 
near $k=0$, 
and therefore short-ranged in $k$ space. 
Thus we may perform a gradient expansion
for $\Delta({\bf R}, {\bf k})$ in $k$ space, as in previous sections.
Introducing ${\bf k}=({\bf k}_1+{\bf k}_2)/2$, ${\bf k}'={\bf k}_2-{\bf k}_1$,
we obtain
\beqarr
&{\beta\over A^2}&\int{d{\bf R}}\sum_{{\bf k}_1{\bf k}_2}
\Delta({\bf R}, {\bf k}_1)\overline{\Delta}({\bf R}, {\bf k}_2)
F(|{\bf k}_1-{\bf k}_2|)\nonumber\\
&=&{\beta\over A^2}\int{d{\bf R}}\sum_{{\bf k}{\bf k}'}
(|\Delta({\bf R}, {\bf k})|^2
-{|{\bf k}'|^2\over 4}
|\nabla_{\bf k}\Delta({\bf R}, {\bf k})|^2+ \cdots) F(k')
\nonumber\\
&=&{\beta\over A}\sum_{\bf k}\int{d{\bf R}}[B_1|\Delta({\bf R}, {\bf k})|^2
+B_2|\nabla_{\bf k}\Delta({\bf R}, {\bf k})|^2+ \cdots],
\eeqarr
where
\beqarr
B_1&=&{1\over A}\sum_{\bf k}F(k)
={1\over V(0)},\\
B_2&=&-{1\over A}\sum_{\bf k}F(k){k^2\over 4}
={1\over 4}\nabla^2({1\over V(r)})|_{r=0}.
\eeqarr
For $V(r)=V_0e^{-r^2/2L^2}$, we have 
\beqarr
B_1&=&{1\over V_0},\\ 
B_2&=&{1 \over 2V_0L^2}.
\eeqarr
Clearly, the coefficient $B_2$ controls the fluctuations of $\Delta$ 
in the ${\bf k}$ space, which describes the internal degrees of freedom
for the pairing amplitude; the larger the interaction range $L$ is, the softer
such fluctuations are. On the other hand, in the short range limit
$L\rightarrow 0$, such fluctuations get completely suppressed, and the
order parameter $\Delta$ depends only of the center of mass coordinate
${\bf R}$, and we recover the standard Ginzburg-Landau theory in terms of
$\Delta({\bf R})$ with no ${\bf k}$ dependence.\cite{popov} 
It is clear however, we need to keep full ${\bf k}$ dependence of $\Delta$
in the present case.

In the first term of Eq. (\ref{s2}), there is no mechanism to control the
spatial (${\bf R}$) fluctuations of $\Delta$. Such fluctuations are controlled
by the second term, which we now turn to.
The second term in Eq. (\ref{s2}) takes the form
\beqarr
&-&\int d{\bf x}_1d{\bf y}_1d{\bf x}_2d{\bf y}_2Q({\bf x}_1, {\bf y}_1;
{\bf x}_2, {\bf y}_2) \Delta({\bf x}_1, {\bf y}_1)
\overline{\Delta}({\bf x}_2, {\bf y}_2)\nonumber\\
&=&-{1\over A}\sum_{\bf k}\int{d{\bf R}_1d{\bf R}_2}
\Delta({\bf R}_1, {\bf k})\overline{\Delta}({\bf R}_2, {\bf k})
P({\bf R}_2-{\bf R}_1, {\bf k}),
\label{s22}
\eeqarr
where
\beq
P({\bf R}_2-{\bf R}_1, {\bf k})={1\over A}\sum_{{\bf k}_1, i\omega_n}
G_0({\bf k}_1, i\omega_n)G_0(2{\bf k}+{\bf k}_1, -i\omega_n)
e^{i(2{\bf k}+2{\bf k}_1)\cdot({\bf R}_2-{\bf R}_1)},
\eeq
and
\beq
G_0({\bf k}, i\omega_n)={1\over i\omega_n-\epsilon_{\bf k}}.
\eeq
In this term there is coupling between $\Delta$'s with different ${\bf R}$'s,
but no coupling between $\Delta$'s with different ${\bf k}$'s.
Thus the first and second quadratic terms in the action control the 
internal ({\bf k}) and spatial (${\bf R}$) fluctuations of the order
parameter $\Delta$ respectively.

The fluctuations of the overall magnitude of the order parameter are controlled
by higher order terms in the effective action. As usual we may stop at the
quartic term near $T_c^{MF}$, and neglect terms that involve the spatial gradient
of the order parameter at this order.\cite{popov}
For the present problem, we obtain
(in a way similar to Ref. \onlinecite{popov})
\beq
S_e^{(4)}={1\over 8A}\sum_{\bf k}\int{d{\bf R}}|\Delta({\bf R}, {\bf k})|^4
({\tanh(\beta\epsilon_{\bf k}/2)\over 
\epsilon_{\bf k}^3}-{\beta\over 2\epsilon_{\bf k}^2
\cosh^2(\beta\epsilon_{\bf k}/2)}).
\eeq

Now we perform a spatial
gradient expansion in ${\bf R}$, for the second quadratic
term, Eq. (\ref{s22}). 
To do that, we first go to momentum space, and define
\beq
\Delta({\bf Q}, {\bf k})=\int{d{\bf R}}e^{-i{\bf Q}\cdot{\bf R}}
\Delta({\bf R}, {\bf k}).
\eeq
In terms of $\Delta({\bf Q}, {\bf k})$, the second quadratic term reads
\beq
{1\over A^2}\sum_{{\bf Q}, {\bf k}}D({\bf Q}, {\bf k})
|\Delta({\bf Q}, {\bf k})|^2,
\eeq
where 
\beqarr
D({\bf Q}, {\bf k})&=&\sum_{i\omega_n}G_0({\bf Q}/2-{\bf k}, i\omega_n)
G_0({\bf Q}/2+{\bf k}, -i\omega_n)\nonumber\\
&=&{\beta\over 2}{\tanh(\beta\epsilon_{{\bf Q}/2-{\bf k}}/2)
+\tanh(\beta\epsilon_{{\bf Q}/2+{\bf k}}/2)
\over \epsilon_{{\bf Q}/2-{\bf k}}+\epsilon_{{\bf Q}/2+{\bf k}}}.
\eeqarr
Expanding for small $Q$, we obtain
\beq
D({\bf Q}, {\bf k})\approx D(0, {\bf k})
-{\beta^3v_F^2\sinh(\beta\epsilon_{\bf k}/2)
\over 32\epsilon_{\bf k}\cosh^3(\beta\epsilon_{\bf k}/2)}({\bf Q}\cdot\hat{\bf k})^2.
\eeq
Fourier transforming back to real (${\bf R}$) space, we obtain
\beq
-{1\over A}\sum_{\bf k}
D(0, {\bf k})\int{d{\bf R}}|\Delta({\bf R}, {\bf k})|^2
+{1\over A}\sum_{\bf k}{\beta^3v_F^2\sinh(\beta\epsilon_{\bf k}/2)\over
32\epsilon_{\bf k}\cosh^3(\beta\epsilon_{\bf k}/2)}
\int{d{\bf R}}|\hat{\bf k}\cdot\nabla\Delta({\bf R}, {\bf k})|^2.
\eeq

Thus putting all these terms together, and keeping leading gradient terms 
only, we obtain
\beqarr
S_e[\Delta,\overline{\Delta}]&=&{\beta\over A}\sum_{\bf k}\int{d{\bf R}}\{
[B_1-C(\beta, {\bf k})]|\Delta({\bf R}, {\bf k})|^2
+B_2|\nabla_{\bf k}\Delta({\bf R}, {\bf k})|^2\nonumber\\
&+&{1\over 2m_{\bf k}}|\hat{\bf k}\cdot\nabla\Delta({\bf R}, {\bf k})|^2
+U(k)|\Delta({\bf R}, {\bf k})|^4\},
\label{glaction}
\eeqarr
where 
\beqarr
B_1&=&1/V_0,\\
B_2&=&1/(2V_0L^2),\\
C(\beta, k)&=&\tanh{\beta\epsilon_k\over 2}
/(2\epsilon_k),\\
{1\over 2m_{\bf k}}&=&{\beta^2v_F^2\sinh{\beta\epsilon_k\over 2}\over
32\epsilon_k\cosh^3{\beta\epsilon_k\over 2}}, \\ 
U(k)&=&
{\tanh{\beta\epsilon_k\over 2}\over 8\epsilon_k^3}-{\beta\over 16\epsilon_k^2\cosh^2
{\beta\epsilon_k\over 2}}.
\eeqarr
Eq. (\ref{glaction}) is one of the central results of this paper; it is the
basis for the calculation of reduction of $T_c$ from $T_c^{MF}$ due to thermal
fluctuations of collective modes in the next section.

Thus far in our discussion we have assumed a uniform space with no 
background electromagnetic field. In the presence of a weak and slowly
varying background magnetic field, using standard arguments\cite{popov}
one finds that the only modification one needs to make in (\ref{glaction}) is
to replace $\nabla$ by $\nabla+2ie{\bf A}({\bf R})/c$.
The supercurrent is 
\beq
{\bf j}({\bf R})=-c{\delta S_e\over \delta{\bf A}({\bf R})}
={ie\beta\over A}\sum_{\bf k}{{\bf k}\over m_k}\{\overline{\Delta}
[{\bf k}\cdot(\nabla+{2ie{\bf A}\over c})\Delta]-c.c.\}.
\eeq
>From this we may obtain the equivalent of London's equation:
\beq
{\bf j}({\bf R})={-4\beta 
e^2\over Ac}\sum_{\bf k}|\Delta_{\bf k}|^2{{\bf k}\over m_k}
[{\bf k}\cdot {\bf A}({\bf R})]=\{{-4e^2\over Acd}
\sum_{\bf k}{|\Delta_{\bf k}|^2k^2\over
m_k}\}{\bf A}({\bf R}),
\eeq
from which the expression for penetration depth follows:
\beq
{1\over \lambda^2}
={8\beta e^2\over Adc^2}\sum_{\bf k}{|\Delta_{\bf k}|^2k^2\over
m_k}.
\eeq
Here $d$ is the spacing between layers. 
These are, of course, mean-field results;
fluctuations have been left out at this level.
And they apply only near $T_c^{MF}$, even at the mean-field level.

\section{reduction of $T_c$ due to thermal
fluctuations of collective modes and pseudogap behavior}
\label{tcreduction}

We study in this section the reduction of of $T_c$ from $T_c^{MF}$ due to
the fluctuations of the collective modes, using the Ginzburg-Landau free
energy derived earlier. As a warm-up, as well as for the purpose of 
later
comparison, we study the same effect in a weak coupling BCS superconductor
with short range ($\delta$-function) interaction, whose Ginzburg-Landau free
energy is the familiar $O(N)$ $\phi^4$ theory with $N=2$:\cite{popov}
\beq
S_e=\beta\int{d{\bf r}}\left(-{1\over 4m_e}|\nabla\psi|^2+a|\psi|^2
+b|\psi|^4\right),
\eeq
where in 3D we have
\beqarr
\psi({\bf r})&=&
\left({7\zeta(3)m_ek_F\epsilon_F\over 12 \pi^4(T_c^{MF})^2}\right)
^{1/2}\Delta({\bf r}),\\
a&=&{6\pi^2T_c^{MF}(T-T_c^{MF})\over 7\zeta(3)\epsilon_F},\\
b&=&{9\pi^4(T_c^{MF})^2\over  14\zeta(3)m_ek_F\epsilon_F^2},
\label{phi4}
\eeqarr
and $m_e$ is the effective mass of the electron.

In the mean field theory, one neglects the quartic term in (\ref{phi4}), and
$T_c$ is determined by the point where $a$ turns negative, which is nothing
but $T_c^{MF}$. The fluctuation effects of the quartic term may be studied
using the self-consistent field approximation that is exact in the limit
$N\rightarrow\infty$.\cite{chaikin}
Within this approximation, one replaces
$|\psi|^4$ by $4\langle|\psi|^2\rangle|\psi|^2$, and requires that the
following self-consistency equation is satisfied:
\beq
\langle|\psi|^2\rangle={1\over V}
\sum_{\bf k}\langle\psi_{\bf k}\psi_{\bf -k}\rangle
={1\over (2\pi)^3}\int{d^3{\bf k}}{k_BT\over {k^2\over 4m_e} + a+
4b\langle|\psi|^2\rangle}.
\eeq
At $T=T_c$, we ought to have $a+
4b\langle|\psi|^2\rangle=0$. Thus the equation that determines $T_c$ becomes
\beq
a+4b{4m_ek_BT_c\over (2\pi)^3}
\int_{k<\Lambda_{uv}}{d^3{\bf k}\over k^2}=0,
\eeq
where $\Lambda_{uv}\sim \hbar\omega_D/v_F$ is an appropriate high 
momentum (ultraviolet) cutoff for $k$. From this we obtain
\beq
{T_c^{MF}-T_c\over T_c^{MF}}\sim {k_BT_c\hbar\omega_D\over \epsilon_F^2}
\ll 1
\eeq
for weak coupling superconductors, whence it is clear that the mean field value for 
$T_c$ is extremely accurate.

In the following we demonstrate that the situation is very different, and
$T_c$ becomes significantly lower than $T_c^{MF}$ when $L\gg \xi$, due to 
thermal fluctuations of the low energy collective modes discussed in previous
sections. We use the Ginzburg-Landau theory appropriate for this situation,
Eq. (\ref{glaction}), and use the same self-consistent field approximation as
above by replacing $|\Delta({\bf R}, {\bf k})|^4$ with
$4\langle|\Delta({\bf R}, {\bf k})|^2\rangle|\Delta({\bf R}, {\bf k})|^2$,
after which the reduced action takes a quadratic form:
\beq
S_R[\Delta,\overline{\Delta}]={\beta\over A}\sum_{\bf k}\int{d{\bf R}}\left\{
[B_1-C(\beta, {\bf k})+\tilde{U}(k)]|\Delta({\bf R}, {\bf k})|^2
+B_2|\nabla_{\bf k}\Delta({\bf R}, {\bf k})|^2
+{1\over 2m_{\bf k}}|\hat{\bf k}\cdot\nabla\Delta({\bf R}, {\bf k})|^2\right\},
\eeq
where
\beq
\tilde{U}(k)=4U(k)\langle|\Delta({\bf R}, {\bf k})|^2\rangle.
\eeq
We would like to formally diagonalise this quadratic form. To do that, we
take advantage of the translation invariance of the action, and 
express the reduced action in terms of the Fourier transform of
$\Delta({\bf R}, {\bf k})$, $\Delta({\bf Q}, {\bf k})$:
\beq
S_R[\Delta,\overline{\Delta}]={\beta\over A^2}
\sum_{{\bf k},{\bf Q}}\left\{\left[B_1-C(\beta, {\bf k})+\tilde{U}(k)+
{(\hat{\bf k}\cdot{\bf Q})^2\over 2m_k}\right]|\Delta({\bf Q}, {\bf k})|^2
+B_2|\nabla_{\bf k}\Delta({\bf Q}, {\bf k})|^2\right\}.
\eeq
Note that modes with different ${\bf Q}$'s decouple.
To proceed further, we introduce eigen modes $\psi_{mn}({\bf Q}, {\bf k})$
which satisfy
\beq
\left[B_1-C(\beta, {\bf k})+\tilde{U}(k)+
{(\hat{\bf k}\cdot{\bf Q})^2\over 2m_k}-B_2\nabla_{\bf k}^2\right]
\psi_{mn}({\bf Q}, {\bf k})=E_{mn}(Q)\psi_{mn}({\bf Q}, {\bf k}),
\label{eigen}
\eeq
and the normalization condition:
\beq
{1\over A}\sum_{{\bf k}}|\psi_{mn}({\bf Q}, {\bf k})|^2=1.
\eeq
Here $m$ and $n$ are quantum numbers to be specified later.
Expanding $\Delta({\bf Q}, {\bf k})$ in terms of $\psi_{mn}({\bf Q}, {\bf k})$:
\beq
\Delta({\bf Q}, {\bf k})=\sum_{mn}a_{mn}({\bf Q})\psi_{mn}({\bf Q}, {\bf k}),
\eeq
we bring the reduced action to diagonal form:
\beq
S_R[\Delta,\overline{\Delta}]={\beta\over A}\sum_{{\bf Q}mn}
E_{mn}(Q)|a_{mn}({\bf Q})|^2.
\eeq
The self-consistent equation now becomes
\beq
\tilde{U}(k)=4U(k)\langle|\Delta({\bf R}, {\bf k})|^2\rangle
={4U(k)\over A^2}\sum_{{\bf Q}mn}|\psi_{mn}({\bf Q}, {\bf k})|^2
\langle|a_{mn}({\bf Q})|^2\rangle
={4U(k)\over A\beta}\sum_{{\bf Q}mn}{|\psi_{mn}({\bf Q}, {\bf k})|^2\over
E_{mn}(Q)}.
\label{selfconsistent}
\eeq
At $T=T_c$, we have the lowest eigenvalue
\beq
E_{00}(Q=0)=0;
\label{tc}
\eeq
This poses a self-consistent condition that determines $T_c$.

The Schroedinger-like equation (\ref{eigen}) is quite difficult to solve in
general, one of the reason being the isotropy in ${\bf k}$ space is lost due
to the presence of the term proportional to $(\hat{\bf k}\cdot{\bf Q})^2$. In 
order to 
proceed, we need to make a number of further approximations and
assumptions. Firstly, we average $(\hat{\bf k}\cdot{\bf Q})^2$ along different 
directions of ${\bf k}$ and replace it by $Q^2/2$, so that isotropy in
${\bf k}$ space is restored and all modes may be labeled by an 
``angular momentum" quantum number $m$. We also notice that $-C(\beta, k)$
has its minimum at $k=k_F$, and goes to zero rapidly for 
$\epsilon_k > 1/\beta$. 
Hence we may approximate it by a ``square'' well, with depth $-C(\beta, k_F)
= {\beta\over 4}$ and
width ${2\over v_F\beta}$.
$U(k)$ (and therefore $\tilde{U}(k)$)
has a similar structure, and we can also approximate it by a step-like
structure, with height $U_{k_F}={\beta^3\over 96}$ and the same
width ${2\over v_F\beta}$.
Anticipating that the ``low-energy" ($E$) modes will be localized in the 
step well, we may approximate
${1\over 2m_k}$ by a constant ${1\over 2m_{k_F}}$.
With these simplifications, 
Eq. (\ref{eigen}) reduces to that of the Schroedinger equation of a particle
confined to a step potential well. Imposing periodic boundary condition at
the ends of the well,\cite{note3}
we obtain
\beq
\psi_{mn}({\bf Q}, {\bf k})=\sqrt{{\pi v_F\beta\over k_F}}
e^{i[2\pi m\theta_{\bf k}+n\pi(k-k_F)/v_F\beta]},
\eeq
and at $T=T_c$ (using (\ref{tc}))
\beq
E_{mn}(Q)={\beta^3v_F^2Q^2\over 64}+B_2({m^2\over k_F^2}+\pi^2n^2v_F^2\beta^2).
\label{spec}
\eeq

Combining Eqs. (\ref{selfconsistent}) and (\ref{spec}), we obtain the equation
that determines $T_c$:
\beq
B_1-{\beta_c\over 4}+{2\over 3\pi k_Fv_F}\sum_{mn}\int
{d{\bf Q}\over Q^2+{64B_2\over \beta_cv_F^2k_F^2}(m^2+\pi^2n^2\beta_c^2
v_F^2k_F^2)}=0.
\eeq
We notice that for the case of $m=0$ and $n=0$, the integral is logarithmically
divergent at both infrared and ultraviolet. As discussed earlier, there is 
always an ultraviolet cutoff $\Lambda_{uv}\approx E_c/v_F$. 
The infrared divergence is 
a signature of the fact that 2D is the lower dimensionality for ordering of 
this model (we do not go into details of the Kosterlitz-Thouless picture for
a true phase transition for $N=2$ here). An infrared cutoff 
$\Lambda_{ir}$ is provided by 
invoking the quasi-2D nature of all real systems, which eventually crossover
to 3D at sufficiently long length scales. Assuming $\Lambda_{ir}$ is
sufficiently large compared to the energies (measured in proper units) of the
modes whose fluctuations contribute significantly to the reduction of $T_c$, 
we obtain
\beq
\beta_c\approx{4M\over 3v_Fk_F}\log{\Lambda_{uv}\over \Lambda_{ir}},
\eeq
where $M$ is the number of modes contributing in the sum of $m$ and $n$.

To determine $M$, we note that the self-consistent potential (for $Q=0$; finite
$Q$ only adds a constant to it) is zero inside the well, and $B_1$ 
outside it. Thus in order for the previous approximations for mode solutions
and ``energies" to be valid, we need to have $E_{mn}(Q=0) < B_1$. Summing
up the number of these modes, we obtain
\beq
M\sim k_F\sqrt{B1/B_2}=k_FL.
\eeq
We thus find
\beq
T_c=k_B/\beta_c\sim{v_F\over L\log{\Lambda_{uv}\over \Lambda_{ir}}}
\sim {T_c^{MF}\over (L/\xi)\log{\Lambda_{uv}\over \Lambda_{ir}}}.
\label{ourtc}
\eeq  
It clearly goes to zero as $L\rightarrow \infty$.\cite{cutoffnote2}
We would like to emphasize that here we have made a number of crude 
approximations in the calculation of $T_c$, thus the dependence of $T_c$ on the
range $L$ may not be quantitatively reliable. However it is quite clear that
the approximations we made tend to {\em underestimate} the importance of the
fluctuation effects; thus the qualitative conclusion that $T_c\rightarrow 0$
as $L\rightarrow\infty$ must hold.

\section{Summary and discussion}
\label{summary}

In the bulk of the paper, we have taken the attitude that the model we 
study here, namely a system with long (compared to the coherence length)
but finite range pairing interaction, 
is a theoretical model that is interesting in its own right, and worked out
some unusual properties of this model, with emphasis on those properties 
that are qualitatively different from those of the standard weak coupling
BCS superconductors stabilized by a short-range pairing potential. Our most
interesting finding is that in this model the transition temperature $T_c$ is
controlled by thermal fluctuations of collective modes, and can be significantly
lower than the quasiparticle gap $\Delta$ or the mean-field transition 
temperature $T_c^{MF}$; as a consequence in the temperature range $T_c < T
 < T_c^{MF}$ the system exhibits pseudogap behavior as the electrons are still
paired while there is no superconducting long-range order.\cite{ghosh}
In this section we
attempt to make contact between our results and the phenomenology of cuprate
superconductors, discuss the relation between our model and existing theoretical
work on the pseudogap behavior, 
point out the limitations of our model as well as 
of our analysis, and indicate some natural extensions and directions 
for future study.

\begin{figure*}[h]
\centerline{\epsfxsize=14cm
\epsfbox{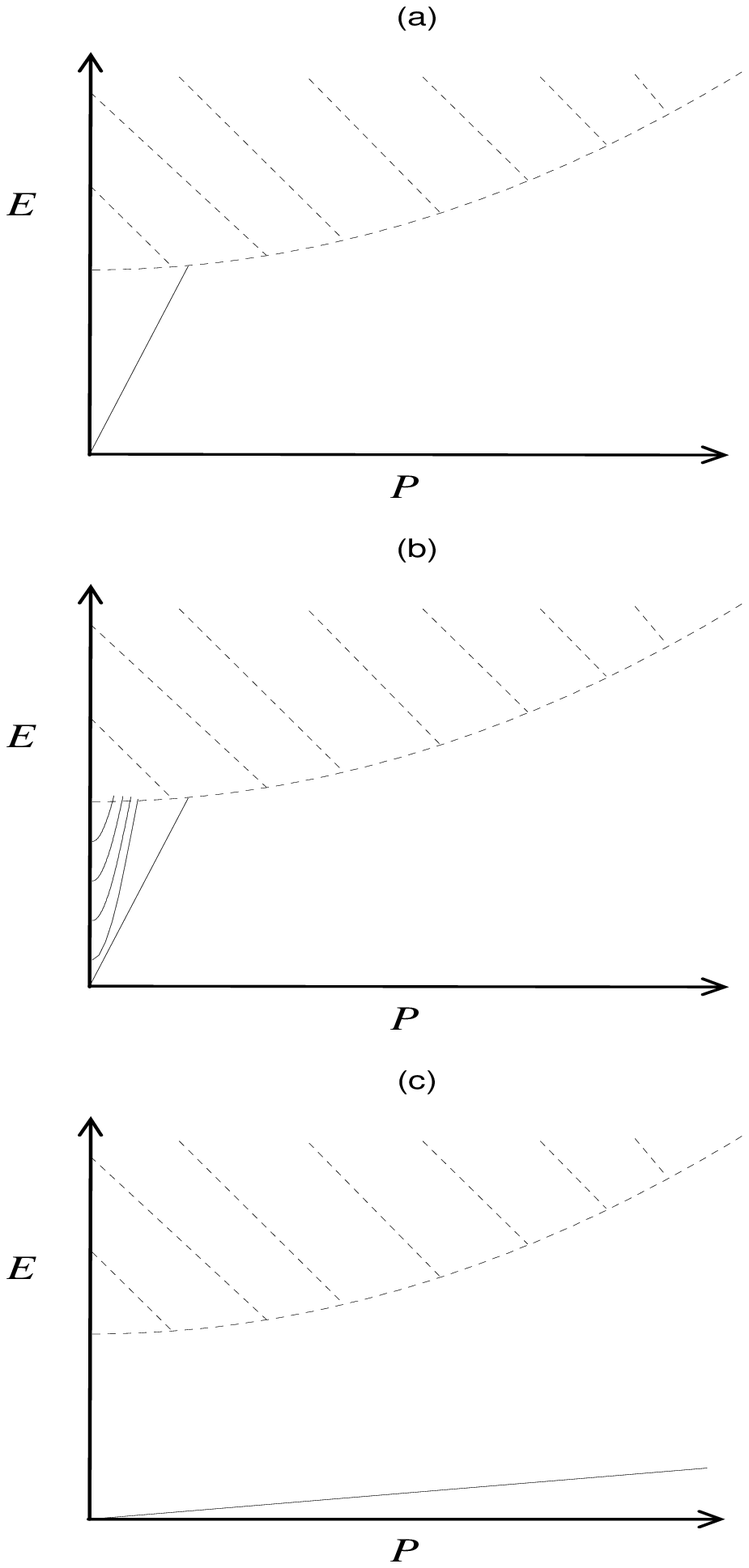}
}
\caption{
Schematic illustration of the neutral
excitation spectra of three different types of
superconductors, in the absence of Coulomb interactions. 
The region shaded by dashed lines with a gap $2\Delta$
is the quasiparticle pair
continuum, common to all three cases; the solid lines stand for collective 
modes. For a weak-coupling BCS superconductor
with short-range pairing interaction (part (a); $\xi\gg 1/k_F$ and
$\xi\gg L$, where $\xi$ is the coherence length, 
$k_F$ is the Fermi wave vector,
and $L$ is the range of the pairing interaction),
the only collective excitation is the
linear Goldstone mode, whose velocity ($v\sim v_F$) is very big as compared to
the scale of the
gap: $vk_F\gg 2\Delta$, thus contributes little to the thermodynamics and
hence the determination of $T_c$. In the BEC case (part (c); $\xi\ll 1/k_F$), 
again the
linear Goldstone mode is the only low-energy
collective mode, but due to the reversed
energy scale: $vk_F\ll 2\Delta$, it dominates the thermodynamics and
sets the scale for $T_c$. The situation studied in this paper, as illustrated
in part (b), is in some sense between these two extremes, namely we have
$\xi\gg 1/k_F$ as in weak-coupling BCS, but in the meantime $\xi\ll L$. 
Here the Goldstone
mode is very steep and hence contributes little to the thermodynamics; however
the the combined effect of all the low-energy modes, stabilized by the 
condition $\xi \ll L$, dominates the thermodynamics and determines $T_c$.
}
\label{fig:spectra}
\end{figure*}

One of the motivations of the present study is the observation that the 
coherence length $\xi$ is much shorter in the cuprates than in conventional 
superconductors. We are, however, by no means the first to 
suggest that a short $\xi$ can lead to behavior qualitatively different from
weak coupling BCS theory. Since the early days of high $T_c$, following the work
of Leggett and
Nozieres and Schmitt-Rink,\cite{noz-sr} 
Randeria and coworkers,\cite{randeria,melo,engelbrecht,randreview} 
as well as others,\cite{levin,tremblay} have argued 
that the short coherence length may bring the cuprates to a regime that is 
intermediate between the weak-coupling BCS limit and the Bose-Einstein 
Condensation (BEC) limit of Cooper pairs. The latter case is realized if 
$\xi$ is much shorter than the inter-particle spacing so that $k_F\xi\gg 1$;
in terms of energy scales, that corresponds to the case $\Delta\gg E_F$.
In this case the Cooper pairs are so closely bound that they hardly overlap, 
and may be viewed as point-like hard-core bosons at low energies, while the
transition temperature $T_c$ is essentially their Bose-Einstein condensation 
temperature which is much lower 
than and unrelated to the pairing energy $\Delta$.
For $T_c < T < \Delta$, the electrons remain paired yet there is no long-range
superfluid order, hence pseudogap behavior. In terms of low-energy excitations
responsible for destroying superconductivity, in the BEC limit it is the
linear Goldstone mode (assuming no Coulomb interaction), while the quasiparticle
excitations (broken pairs) cost too much energy to have any effect on $T_c$.
Put differently, the thermal fluctuations of this Goldstone mode are the
classical phase fluctuations of the superconducting order parameter that control
the transition in this limit. (We should note that Emery and Kivelson\cite{ek}
have argued that classical phase fluctuations {\it are} the physics of the
pseudogap regime in the underdoped materials, but their point of departure is
logically distinct, deriving from a small zero temperature superfluid stiffness
that is connected to the physics of a doped Mott insulator.)
In the cuprates, $k_F\xi\sim 10$, and $\Delta$ is still a small energy compared
to $E_F$ (although the difference is not nearly
as overwhelming as in conventional
superconductors); we are thus still somewhat distant from the BEC
limit. The interesting new feature of the model studied here is
that by having $\xi$ much smaller than the range of pairing interaction $L$,
we can get the pseudogap behavior while staying the {\em weak coupling} regime
(in the sense $k_F\xi\gg 1$ and $\Delta/E_F\ll 1$). In this case the Goldstone
mode is unable to drive $T_c$ below the scale of $\Delta$ by itself; it needs
all the help from the other collective modes supported by this model.
The low-energy spectra of the weak-coupling BCS superconductor, the BEC 
superconductor, and the present model are summarized schematically in 
Fig. \ref{fig:spectra}. While there are clearly qualitative differences 
between the BEC (as well as phase fluctuation) picture and the present model, 
they share the common spirit that $T_c$ is determined by collective modes
instead of quasiparticle excitations. 
It is also worth mentioning that in real
systems with long-range Coulomb interaction, the energy of the Goldstone mode
is expected to be pushed up to the plasmon frequency; the importance of this
fact to the thermal fluctuations of this mode (or classical phase 
fluctuations) is still under discussion.\cite{carlson} On the other hand the
Coulomb interaction is not expected to significantly
affect the spectra of the (gapped) 
exciton-like modes discussed here. This is due to the different nature of
the Goldstone mode and exciton modes: the former is a consequence of
the broken gauge symmetry, and introducing the Coulomb interaction (or, more
generally, the electromagnetic interaction) converts the Goldstone mechanism of
broken symmetry to the Higgs mechanism. The exciton modes, 
on the other hand, are the
bound states formed by quasiparticle pairs
due to the residual attractive interaction 
between quasiparticles not included in the mean field approximation.
We therefore believe the presence of 
Coulomb interaction does not affect the fluctuation physics discussed here
significantly.

As pointed out already in the Introduction, in some 
theoretical models for cuprate superconductivity,
the range of the interaction that gives rise to Cooper
pairing can be very long. In the interlayer pair hopping 
model,\cite{anderson,chakravarty} it is assumed that pairing is induced by a 
pair hopping term in the Hamiltonian, that is diagonal in momentum space. 
Fourier-transforming to real space, this corresponds to an {\em infinite}
range hopping term for Cooper pairs, corresponding (loosely)
to the $L\rightarrow\infty$
limit of the model we study here (the fact that the hopping term is off-diagonal
in layer index is of no qualitative consequence). In that {\em specific form},
it has already been shown\cite{sarker} that the model may be solved exactly
in the absence of any other in-plane paring interaction, the transition
temperature is zero, and there is pseudogap behavior at low temperatures.  
Our results are in agreement with this observation. The model we use here, 
however, is
more general and versatile, and in particular enables one to address how the
pseudogap behavior develops as the range $L$ increases, and how $T_c$ 
approaches zero as $L\rightarrow\infty$.

It is equally interesting to scrutinize the results obtained here in the
context of the spin fluctuation theory\cite{pines} of cuprate superconductivity.
In this model the range of the interaction $L$ is essentially
the spin-spin correlation 
length.\cite{lengthnote}
Thus $L$ is of order lattice spacing in the overdoped region of the
phase diagram, increases as the doping level $x$ decreases, becomes much longer 
than the lattice spacing in the underdoped region, 
and eventually diverges upon approaching the 
antiferromagnetic phase boundary at very low doping. The pseudogap behavior 
is observed in the underdoped regime, where $T_c\rightarrow 0$ while the gap
(the maximum of the $d$-wave gap measured at very low $T$ by, say,
photoemission)
slowly {\em increases} as $x$ decreases. This behavior is certainly 
consistent with our findings here; the reduction of $T_c$ and its departure 
from the gap is due to the increase of $L$, while the size of gap,
which in 
our model saturates at the depth of the potential well $V_0$ for large $L$,
would be set by the scale of the near neighbor spin coupling strength $J$.
Hence at a very crude level, we can qualitatively account for the phenomenology
of the underdoped cuprates by combining our results 
with the spin-fluctuation model.

On the experimental side, some circumstantial evidence for both the importance
of the longer range part (beyond near-neighbor) of the pairing interaction, 
and the possible relevance of our results, exists.
i) Recent photoemission measurements\cite{mesot} of gap anisotropy have found
deviations from the standard 
\beq
\Delta_{\bf k}\propto \cos k_x-\cos k_y
\label{dx2y2}
\eeq
dependence
of the $d_{x^2-y^2}$ order parameter in the underdoped region, with the 
deviation increasing with the decreasing doping level. This deviation is 
interpreted\cite{mesot} as due to longer-ranged pairing interaction, as
a nearest-neighbor attraction leads to Eq. (\ref{dx2y2}). 
We consider this to be
(somewhat indirect) evidence that the range of the pairing interaction 
increases with decreasing doping level in underdoped cuprates.
(ii) It has been recently noticed\cite{balatsky} that there is a strong 
correlation between $T_c$ and peak width of the {\em normal state} spin 
susceptibility $\chi(q,\omega)$, in some cuprates. Specifically, 
Balatsky and and Bourges\cite{balatsky} found that in $YBCO_{123}$
and $La_{214}$ compounds, $T_c\propto\delta q$, where $\delta q$ is the width
of $Im \chi(q,\omega)$ at low $\omega$. This lead the authors to conclude
that ``antiferromagnetism is likely responsible for the high $T_c$ 
superconducting mechanism". If so, one is lead to the conclusion that
$T_c\propto 1/L$, where $L\sim 1/\delta q$ is the range of the pairing 
interaction mediated by the {\em normal state} antiferromagnetic spin 
fluctuations, in agreement with our Eq. (\ref{ourtc})! However, given the
facts that our model is greatly oversimplified as far as the cuprates go
(see below) and our treatment of the
fluctuation effects is still preliminary (also below), we 
consider this agreement to be fortuitous at this point. Nevertheless, the
findings of Ref. \onlinecite{balatsky} do indicate the importance of the
range of interaction on $T_c$, and are in qualitative agreement with our
results. 
(iii) Experimentally, it has been found\cite{norman} that upon cooling
underdoped cuprate samples, 
the temperature $T^*$ at which the pseudogap opens up depends on
the location in momentum space; $T^*$ is highest near the superconducting
gap maxima, while 
lowest near the gap nodes; it thus suggests that $T^*$ is a ``local" property
in momentum space. This can be understood quite naturally by invoking a 
long range pairing interaction as studied here, since such an interaction is
localized in momentum space. As we have shown in section \ref{reduced}, one can
define a momentum-dependent mean field transition temperature, 
$T_c^{MF}(\epsilon_{\bf k})$, which increases monotonically 
with $\Delta_{\bf k}$;
it also sets the temperature $T^*\sim T_c^{MF}(\epsilon_{\bf k})$
below which a local gap is opened up and the pseudogap sets in.

We must emphasize that the contact between our model and the cuprate physics
made above is tentative, and in its present form this model can only be viewed
as an interesting toy model that gives rise to pseudogap behavior. In the 
following we discuss some limitations of the model as well as our treatment, 
and some natural directions for extensions.

The ground state of our model is a fully-gapped
$s$-wave superconductor, while the 
cuprates (at least most of them) are known to have a $d$-wave order parameter.
The most important difference between the two is that the latter supports
{\em gapless} nodal quasiparticles. It has already been suggested\cite{lee}
that the thermally excited quasiparticles may be responsible for emptying
the superfluid stiffness, setting the scale of $T_c$, and giving rise to
pseudogap behavior in underdoped cuprates. This important piece of physics is
missing in our model. On the other hand, as long as collective modes of the
{\em order parameter} are concerned, our method can be generalized to $d$-wave
(or other unconventional superconductors) fairly easily. The
key feature of a long (but finite) range pairing interaction is its {\em 
locality} in momentum space, i.e., $V_{\bf k}$ is sharply peaked in ${\bf k}$
space.
This allows a gradient expansion in 
momentum space for the order parameter. In an $s$-wave superconductor, 
$V_{\bf k}$ is sharply peaked at ${\bf k}=0$, while for a $d$-wave 
superconductor appropriate for cuprates, one would choose a $V_{\bf k}$ that
is sharply peaked at a wave vector ${\bf Q}$ which is at
or near $(\pi, \pi)$. This, however, is not going to make
any qualitative difference to the analysis made in this paper.
At a more detailed level, two cases need to be distinguished from each other.
(i) There is Fermi surface nesting and ${\bf Q}$ is exactly or very close to
the nesting wave vector. In this case the entire Fermi surface participates in
pairing actively and the analysis carried out here, based on the
gradient expansion of
the superconducting order parameter both along and perpendicular to the Fermi
surface, carries through straightforwardly. (ii) There is no Fermi surface 
nesting, or ${\bf Q}$ is not close to the nesting wave vector. In this case
only certain ``hot spots" at the Fermi surfaces that are connected by
${\bf Q}$ participate in the pairing actively. In this case one can still 
develop a gradient expansion within the hot spots. In either case the 
qualitative features of our results are expected to be robust.
We note that in a recent study of the spin-fermion-hot spot model,\cite{abanov}
it has been  conjectured
that in the limit of infinite spin-spin correlation
length, the transition temperature obtained from
solving the Eliashberg equation is only the onset of pseudogap behavior, while
the real $T_c$ vanishes in that limit. This 
conjecture is in agreement with our results,
as the spin-spin correlation
length corresponds to the range of interaction $L$ in our model. In the
present paper, we have developed a systematic 
way to study the physics of pseudogap behavior 
due to large $L$, and shown {\em explicitly} that 
$T_c\rightarrow 0$ as $L\rightarrow\infty$.

Our analysis of the thermal fluctuations of the collective modes is based on
a Ginzburg-Landau free energy functional, 
which we derive using a functional integration
formalism and an expansion in power series of the magnitude of the order
parameter. Strictly speaking this power series expansion is valid only in the
vicinity of $T_c^{MF}$, where the amplitude of the order parameter just starts
to develop; at $T=T_c \ll T_c^{MF}$, the amplitude of the order parameter is
big and such an expansion is no longer appropriate. What is also missing are
the contributions from the fluctuations of components of the order parameter
with finite Matsubara frequencies; we neglected them on the ground that we are
primary interested effects of thermal fluctuations; however at low $T$ these
components are important to the physics, especially if one is also interested
in the quantum fluctuations of the order parameter. 
Even with these simplifications, the resultant free energy functional is quite
complicated, and we need to introduce approximations in 
the one-loop calculation of $T_c$, like neglecting the anisotropy in the
internal (relative) space of the order parameter when the center of mass 
carries a finite momentum. It is quite possible that one can do a better job
in analyzing the fluctuation, 
especially in the limit of $L/\xi\rightarrow \infty$.
It is worth noting, however, 
that all our approximations tend to {\em underestimate}
the effects of fluctuations, and thus our basic conclusion that  
$T_c\ll T_c^{MF}\sim \Delta$ when $L/\xi\gg 1$, is clearly valid.

Despite the various disclaimers made above, we hope the model and the crossover
introduced here will serve as a different paradigm of a non-BCS transition to
a superconducting state, which can exhibit pseudogap behavior even without
leaving the weak coupling regime. Its relevance to cuprate physics
is not completely
clear at present, but there are encouraging signs that it is worth 
further pursuit.

\acknowledgements
We have benefited greatly from discussions with and comments from Dan Agterberg,
Nick Bonesteel, Elbio Dagotto, Lev Gor'kov, Peter Littlewood,
and Doug Scalapino.
This work was 
supported by NSF DMR-9971541, the Fairchild and Sloan Foundations
(KY), and NSF DMR-9978074, and the Sloan and Packard Foundations (SLS).

\newpage

\begin{references}
\bibitem{2Dnote} Since our study is motivated by the cuprate 
superconductors which are layered compounds, 
throughout the paper we mostly assume that we are dealing with a 2D system. 
However in our analysis of the fluctuations in Section \ref{tcreduction}
we will invoke the quasi two-dimensional nature of the system to regulate the
``Mermin-Wagner'' infrared singularities by introducing an infrared cutoff on
the 2D momenta.
\bibitem{cutoffnote} Here we assume $E_c \gg V_0$. Of course, according to
Eq. \ref{ECL} $E_c$ will eventually become smaller than $V_0$ when $L$ 
becomes very large. We are not interested in that regime here.
\bibitem{note2} In the subspace where 
the ground 
state belongs to, quasiparticles can only be created in forms of such pairs.
\bibitem{spinnote} If the interactions between electrons with same spin are
kept, one needs to introduce two more sets of
Hubbard-Stratonovich fields (see later),
one for
up-spin pairs and one for down-spin pairs, to decouple these interactions. This
will add some more technical complexity to the analysis, but does not change 
the essential physics.
\bibitem{note3} It is clear that a more appropriate boundary condition is to
require the modes to vanish at the boundaries. However the periodic
boundary condition has the advantage that the mode amplitude is a constant 
inside the well, thus $\tilde{U}(k)$ is also a constant inside the well,
consistent with our assumption.
\bibitem{cutoffnote2} This is the only place the cutoff $E_c$ explicitly
enters a
physical quantity (through $\Lambda_{uv}$), $T_c$, in this paper. 
However the dependence
is so weak (logarithmic), that the dependence of 
$E_c$ on $L$ hardly affects anything.
\end{references}
\end{document}